\documentclass{article}

\PassOptionsToPackage{numbers, compress}{natbib}

 \usepackage[preprint]{neurips_2025}

\usepackage[utf8]{inputenc} 
\usepackage[T1]{fontenc}    
\usepackage{hyperref}       
\usepackage{url}            
\usepackage{booktabs}       
\usepackage{amsfonts}       
\usepackage{nicefrac}       
\usepackage{microtype}      
\usepackage{xcolor}         
\usepackage{graphicx}       
\usepackage{xspace}
\usepackage{multirow}

\newcommand{\kt}{\ensuremath{k_{\mathrm{T}}}\xspace}
\newcommand{\pt}{\ensuremath{p_{\mathrm{T}}}\xspace}

\newcommand{\jetnet}{{\textsc{JetNet}}\xspace}
\newcommand{\jetclass}{{\textsc{JetClass}}\xspace}

\title{RINO: Renormalization Group Invariance with No Labels}

\author{%
  Zichun Hao, Raghav Kansal\thanks{Also affiliated with the Fermi National Accelerator Laboratory, Batavia, IL 60510, USA}\thanks{Now at Bexorg, Inc.}, Chang Sun, Maria Spiropulu
  \\
  Division of Physics, Mathematics and Astronomy \\
  California Institute of Technology \\
  Pasadena, CA 91125 \\
  \texttt{\{zhao,rkansal,chsun,smaria\}@caltech.edu} \\
  \\
  \And
  Abhijith Gandrakota, Jennifer Ngadiuba \\
  Particle Physics Division \\
  Fermi National Accelerator Laboratory \\
  Batavia, IL 60510 \\
  \texttt{\{abhijith,ngadiuba\}@fnal.gov} \\
  \And
  Javier Duarte \\
  Department of Physics \\
  University of California San Diego \\
  La Jolla, CA 92093 \\
  \texttt{jduarte@ucsd.edu}
}

\begin{document}

\maketitle

\begin{abstract}
    A common challenge with supervised machine learning (ML) in high energy physics (HEP) is the reliance on simulations for labeled data, which can often mismodel the underlying collision or detector response.
    To help mitigate this problem of domain shift, 
    we propose \textbf{R}enormalization group \textbf{I}nvariance with \textbf{NO} labels (RINO), a self-supervised learning approach that can instead pretrain models directly on collision data, learning embeddings invariant to renormalization group flow scales. 
    In this work, we pretrain a transformer-based model on jets originating from quantum chromodynamic (QCD) interactions from the \jetclass dataset, emulating real QCD-dominated experimental data, and then finetune on the \jetnet dataset --- emulating simulations --- for the task of identifying jets originating from top quark decays.
    RINO demonstrates improved generalization from the \jetnet training data to \jetclass data compared to supervised training on \jetnet from scratch, demonstrating the potential for RINO pretraining on real collision data followed by fine-tuning on small, high-quality MC datasets, to improve the robustness of ML models in HEP.
\end{abstract}

\section{Introduction} \label{sec:introduction}
Machine learning (ML) applications in high-energy physics (HEP) often face the challenge of the domain shift between Monte Carlo (MC) simulations used for supervised training and real experimental data. 
This mismatch stems from imperfect modeling of detector effects and the underlying physical processes in simulations, and thus can require complicated calibration procedures to ensure reliable physics analyses.
Meanwhile, real collision data are rarely, if ever, used to train ML models because of a lack of labels. 

Self-supervised learning (SSL) offers a compelling avenue to take advantage of experimental data in ML, by training models to learn meaningful representations of the data without labels that have been shown to generalize across tasks and even domains~\cite{SSL-robustness,SSL-single-image,SSL-survey}. 
We therefore explore the potential of SSL in exploiting the vast quantities of unlabeled collision data available for pretraining, followed by supervised fine-tuning on smaller, task-specific MC datasets.
Specifically, we propose \textbf{R}enormalization group \textbf{I}nvariance with \textbf{NO} labels (RINO), an SSL technique similar to DINO~\cite{DINO,DINOv2,DINOv3}, that uses self-distillation to encourage representations invariant to the energy scale of the physical process. 

To test this strategy, we consider datasets of \textit{jets}, which are collimated sprays of particles resulting from the showering and hadronization of quarks and gluon produced at high energy colliders, such as the CERN Large Hadron Collider (LHC)~\cite{Andersson:1987pr}.
Jets are ubiquitous at the LHC and identifying the particles that initiate them is a critical step in data analysis.
They are also high-dimensional, complex data structures --- with 100s of particles per jet, each with multiple features --- providing a fertile playground for ML techniques in physics.
Indeed, jet classification, or ``tagging'', has been the focus of numerous advances in ML~\cite{EFN,interaction-networks,LorentzNet,ParT,PELICAN,L-GATr,Lorentz-equivariant-transformer-LHC,hepmllivingreview}.
Thus, we demonstrate the effectiveness of RINO in reducing domain bias and improving the performance in real data on the downstream task of jet tagging.
Our code is provided in \url{https://github.com/zichunhao/RINO}.

\section{Related Work} \label{sec:related-work}
\paragraph{SSL in HEP}
Recently, there has been growing interest in applying SSL techniques to HEP, including JetCLR~\cite{JetCLR,Zhao:2024kry}, masked particle modeling (MPM)~\cite{MPM,MPM-improved}, resimulation~\cite{Resimulation}, OmniJet-$\alpha$~\cite{OmniJet-alpha}, J-JEPA~\cite{J-JEPA}, and HEP-JEPA~\cite{HEP-JEPA}.
Notably, MPM explores pretraining on real collision data and fine-tuning on simulated datasets, but results show fine-tuned models do not consistently outperform fixed backbone representations, indicating domain adaptation challenges.
Other methods are based on supervised learning for pretraining~\cite{Li:2024htp,Mikuni:2024qsr,Mikuni:2025tar,Mokhtar:2025zqs}.

\paragraph{DINO}
Contrastive learning methods such as SimCLR~\cite{SimCLR}, MoCo~\cite{MoCo}, and SwAV~\cite{SwAV} learn representations by maximizing agreement between augmented views. 
DINO~\cite{DINO,DINOv2,DINOv3} extends this with teacher-student self-distillation, successfully learning attention maps in vision transformers without labels. 
However, standard augmentations in computer vision, such as color jittering and cropping, do not respect the inductive biases of HEP data; for example, unlike images, jet constituents have highly global correlations, with cropping thus an ill-suited augmentation.
In this work, we instead design a physics-informed augmentation for jets. 

\section{Methodology} \label{sec:method}
\subsection{RINO}
RINO is a self-supervised learning framework similar to DINO~\cite{DINO,DINOv2,DINOv3}, where the different ``views'' correspond to its composition at different energy scales during its evolution.
While we cannot directly access this information in data, where we see only its final form as stable hadrons in the detector, we are able to approximate it using its clustering history via the \kt algorithm~\cite{kt-clustering}.
The \kt clustering algorithm is an iterative jet clustering method that prioritizes the combination of objects with low transverse momentum \pt and thus mimics the reverse process of parton showering.
In particular, subjets defined by the \kt algorithm have been shown to correspond well to the underlying hard decay processes~\cite{Thaler:2010tr, CMS:2025eyd}.

From a theoretical perspective, each clustering step integrates out degrees of freedom below a characteristic energy scale. 
Different clustering depths thus probe jets at different scales, with fewer clusters, or subjets, representing coarse-grained descriptions and more subjets preserving fine structure, effectively accessing the jet at different stages of its evolution according to the quantum chromodynamic (QCD) renormalization group flow~\cite{PhysRevD.105.054012}.
Thus, by creating different views through varying \kt clustering steps, RINO motivates models to learn representations invariant to the energy scale, a potent inductive bias in HEP.

\subsection{Training Strategy}
\begin{figure}[ht]
    \centering
    \includegraphics[width=0.55\linewidth]{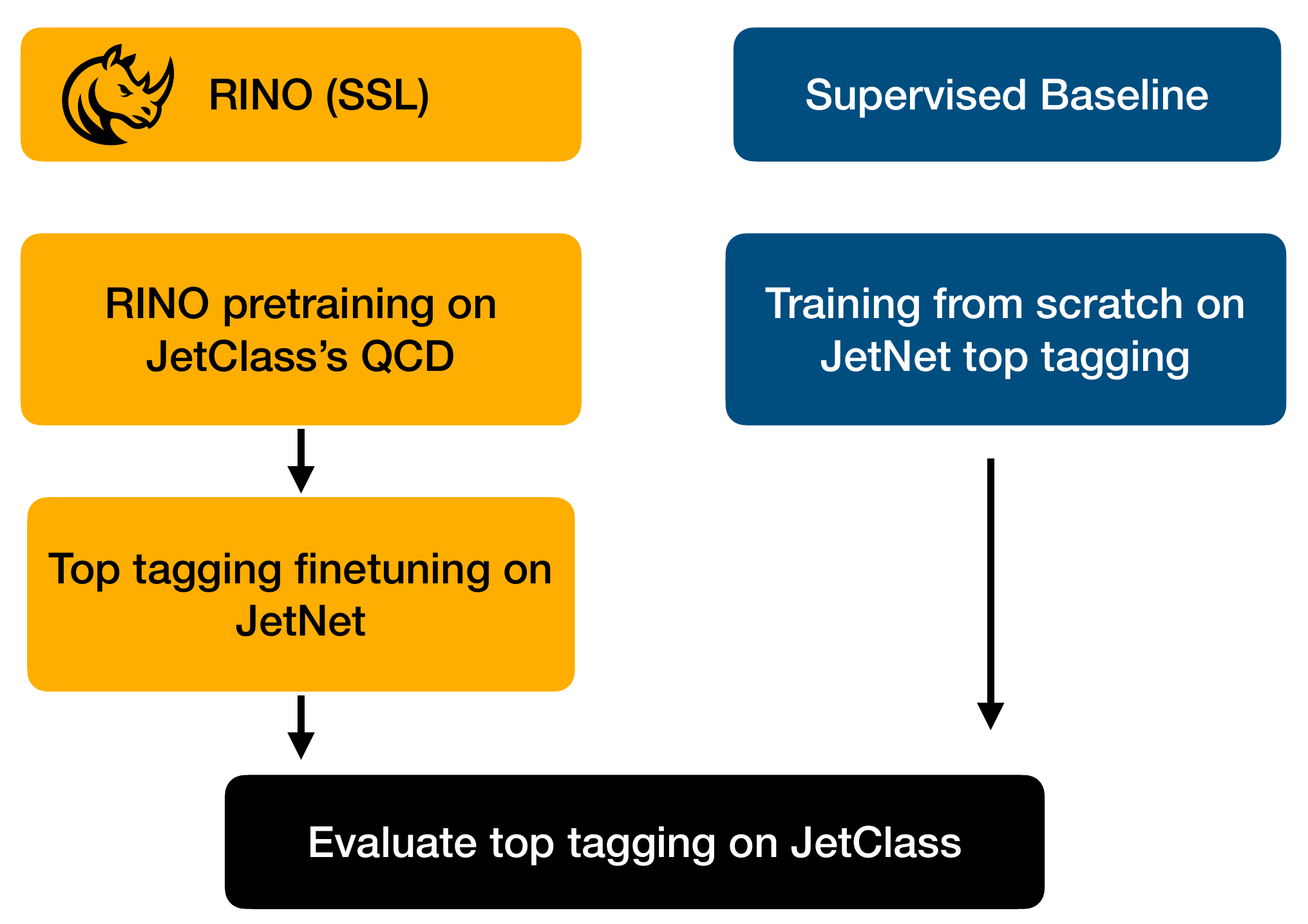}
    \caption{RINO training strategy (left): pretrain on \jetclass QCD jets as a proxy for real data, then finetune on \jetnet top tagging. 
    Supervised baseline strategy (right): train from scratch on \jetnet. 
    Both strategies are then evaluated on top tagging on \jetclass.}
    \label{fig:RINO-workflow}
\end{figure}

To evaluate the efficacy of pretraining on real data in reducing domain bias, we conduct the following experiments, treating the popular \jetnet~\cite{jetnet-dataset} and \jetclass~\cite{jetclass} datasets as simulation and data, respectively.
The different simulation frameworks used in \jetclass and \jetnet allow us to meaningfully approximate the MC-to-data domain shifts commonly encountered in HEP. 

Our baseline strategy is to do a supervised training entirely on \jetnet for the task of discriminating between jets originating from gluons and light quarks (QCD jets) and those from top quark decays --- a task referred to as ``top tagging'' --- and evaluate its performance on \jetclass (Fig.~\ref{fig:RINO-workflow}, right).
This is analogous to the conventional HEP paradigm of training on simulations but applying to data.

The RINO strategy instead proposes first pretraining on data, which is dominated by QCD jets at the LHC.
To emulate this, we pretrain only on QCD jets in \jetclass, then perform a supervised \textit{finetuning} on the smaller \jetnet dataset for top tagging, and finally evaluate them on \jetclass as well.

\section{Experiments} \label{sec:experiment}
\subsection{Models and Training}
We implement four transformer~\cite{transformer} encoder models in PyTorch~\cite{PyTorch} with different sizes and embedding dimensions: nano (32D embedding, 50k parameters), lite (64D, 200k), mini (128D, 1M), and base (256D, 5M).
Full architectural details are provided in Appendix~\ref{appendix:model-details}.

\paragraph{Pretraining}
RINO follows the DINO framework with a teacher-student architecture processing different augmented views of the same jet. 
We use \kt clustering via \textsc{FastJet}~\cite{FastJet} to generate ``global'' views (corresponding to $\{1, 2, 3, 4\}$ subjets) for both networks, while the student additionally receives ``local'' views ($\{8, 16, 32, 64\}$ subjets, along with the original particle-level view). 
The teacher is updated via exponential momentum average (EMA)~\cite{EMA} with momentum starting at $0.992$ and cosine annealing to $1.0$ in the course of training. 

\paragraph{Finetuning}
For the downstream task, we replace the pretraining projection head with task-specific classification heads. 
We explored different adaptation strategies, including linear probing~\cite{alain2018understandingintermediatelayersusing}, which freezes the backbone entirely, and joint finetuning of both head and backbone using reduced learning rate factors ($0.01\times$ and $0.1\times$) for the backbone. 
We observed that linear probing fails to fully leverage the representational capacity of the pretrained backbone, while aggressive finetuning often overfits to the downstream task.
To address these limitations, we adopt the LP-FT (Linear Probing then Fine-Tuning) strategy~\cite{LP-FT, ren2023preparetaskheadfinetuning}, which first trains the classification head with a frozen backbone, and then unfreezes the entire model and continues training with a reduced backbone learning rate.

We evaluate two head architectures: RINO-Linear, which employs a single linear layer directly from the class token representation to the output; and RINO-MLP, which uses a multi-layer perceptron (MLP) with GELU activation and dropout regularization. 
For the supervised baseline strategy, the same model architectures are trained until convergence on the \jetnet dataset. 
Complete training details are given in Appendix~\ref{appendix:training-details}.

\subsection{Results}
\begin{figure}[ht]
   \centering
   \includegraphics[width=0.39\linewidth]{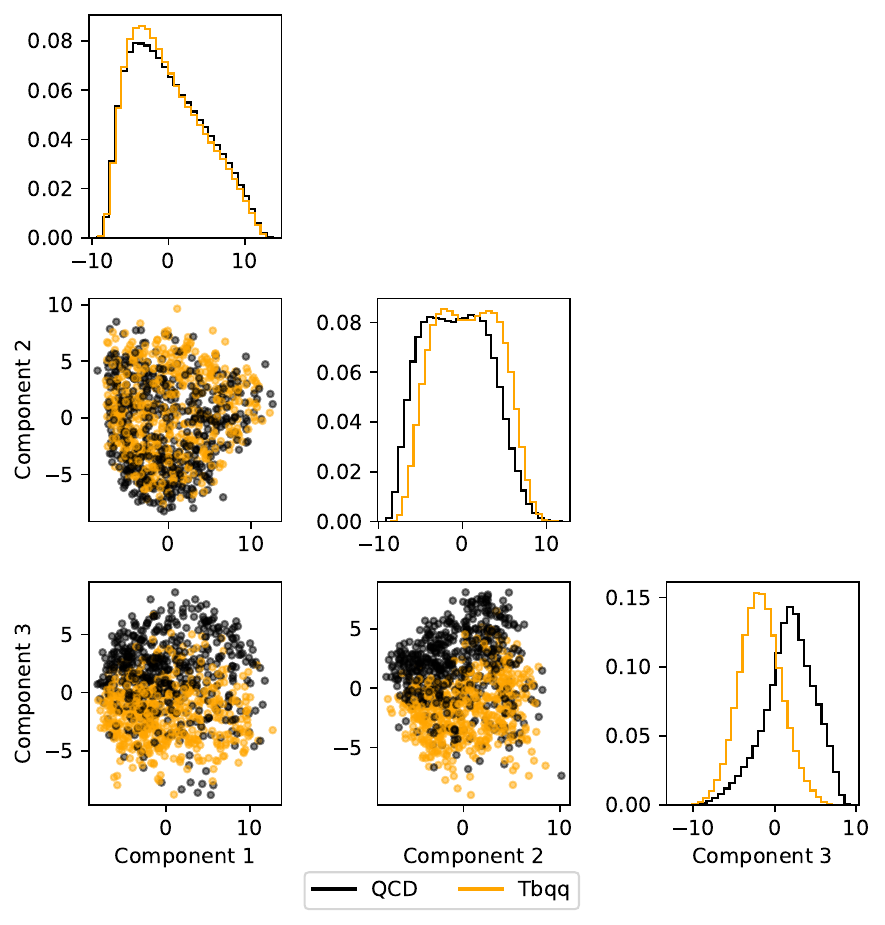}
   \includegraphics[width=0.55\linewidth]{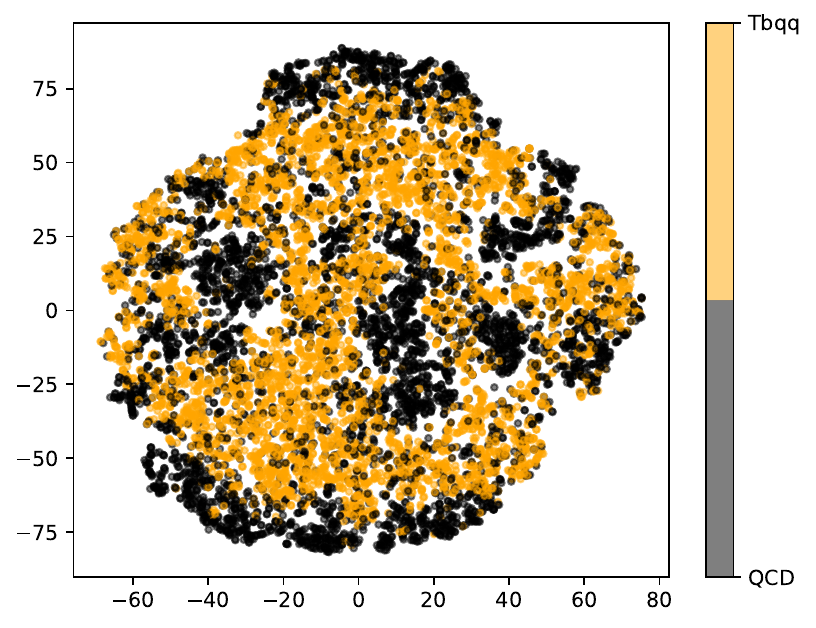}
   \caption{Visualization of learned jet representations from the base model using PCA (left) and t-SNE embedding (right). 
   Jet representations of all five models are shown in Appendix~\ref{appendix:experiments}. 
   }
   \label{fig:embedding-PCA-tSNE-base}
\end{figure}

\paragraph{Backbone Embeddings}
We analyze the pretrained embeddings using \jetclass's hadronic top (\texttt{Tbqq}) class. As shown in Figure~\ref{fig:embedding-PCA-tSNE-base}, both PCA and t-SNE~\cite{t-SNE} reveal good separation of top and QCD jets in the embedding space.
We quantitatively assess embedding quality through $k$-nearest neighbor ($k$-NN) classification~\cite{k-NN} with $k=20$. 
The nano, lite, mini, and base models achieve accuracies of 
$0.866 \pm 0.001$, $0.859 \pm 0.002$, $0.860 \pm 0.001$, and $0.866 \pm 0.002$, respectively. 
To further validate the embedding quality, we evaluate boosted decision trees (BDT)~\cite{BDT} trained on the same embeddings, achieving accuracies of $0.877 \pm 0.002$, $0.879 \pm 0.002$, $0.883 \pm 0.002$, and $0.886 \pm 0.001$ for the respective model sizes.
These results demonstrate that RINO produces rich, discriminative embeddings with meaningful physical information for jet classification, despite being pretrained on QCD jets alone.

\paragraph{Top Tagging}
Table~\ref{tab:accs} summarizes classification performance across model architectures and training strategies on \jetclass and \jetnet. 
All results are obtained using 10-fold cross-validation with 80:20 train-test splits, where different random seeds determine the data partitioning for each fold.
RINO demonstrates improvements over the supervised baseline strategy on \jetclass across all model sizes, with gains ranging from $15\%$ to $26\%$, demonstrating that the RINO pretraining learns transferable representations and reduces the domain bias.
On \jetnet, supervised baselines achieve higher accuracy as expected for in-domain evaluation, with a performance gap ranging from $4\%$ to $5\%$. 
Nevertheless, this is a reasonable trade-off for the substantial cross-domain benefits: RINO maintains competitive performance while providing transferable representations across the two datasets.

\begin{table}[ht]
    \caption{Classification accuracy comparison across model sizes and training strategies on \jetclass and \jetnet datasets. Results report $\mu \pm \sigma$ over 10-fold cross-validation with $80:20$ train-test splits.}
    \label{tab:accs}
    \centering
    \begin{tabular}{c|c|c|c}
       \toprule
       Model size & Strategy & \jetclass Accuracy & \jetnet Accuracy \\ \hline
       nano & Supervised & $0.601 \pm 0.060$ & $\mathbf{0.910 \pm 0.001}$ \\ 
       nano & RINO-Linear & $0.748 \pm 0.005$ & $0.858 \pm 0.001$ \\ 
       nano & RINO-MLP & $\mathbf{0.755 \pm 0.012}$ & $0.863 \pm 0.003$ \\ 
       \hline
       lite & Supervised & $0.551 \pm 0.038$ & $\mathbf{0.910 \pm 0.001}$  \\
       lite & RINO-Linear & $0.810 \pm 0.004$ & $0.864 \pm 0.002$ \\ 
       lite & RINO-MLP & $\mathbf{0.812 \pm 0.005}$ & $0.865 \pm 0.003$ \\ 
       \hline
       mini & Supervised & $0.595 \pm 0.049$ & $\mathbf{0.910 \pm 0.001}$ \\
       mini & RINO-Linear & $\mathbf{0.778 \pm 0.005}$ & $0.857 \pm 0.002$ \\ 
       mini & RINO-MLP & $0.769 \pm 0.013$ & $0.862 \pm 0.002$ \\ 
       \hline
       base & Supervised & $0.629 \pm 0.062$ & $\mathbf{0.910 \pm 0.001}$ \\
       base & RINO-Linear & $\mathbf{0.804 \pm 0.001}$ & $0.868 \pm 0.001$ \\ 
       base & RINO-MLP & $0.772 \pm 0.025$ & $0.869 \pm 0.002$ \\ 
       \bottomrule
    \end{tabular}
\end{table}

\section{Discussion and Broader Impact} \label{sec:discussion}
We introduce RINO, a self-supervised learning framework for learning representations invariant to the energy scale of high-energy physics processes without labels.
Using \jetclass QCD to emulate real collision data for pretraining and then performing supervised finetuning on the \jetnet dataset, emulating simulations, RINO demonstrates significantly improved generalization across domains versus supervised-only baselines, with up to 26\% higher accuracy.
This suggests the potential of pretraining on real experimental data followed by fine-tuning on limited simulations for specific tasks, reducing reliance on MC simulations while improving robustness.
Future work will develop fine-tuning methods to better leverage the rich embeddings learned by RINO.
RINO has the potential for significant broader impact in reducing biases due to simulation-only training in fundamental physics, but future work may also explore more rigorous metrics to quantify robustness.

\begin{ack}
Z.H., R.K., C.S., and M.S. are supported by the California Institute of Technology High Energy Physics under Contract DE-SC0011925 with the United States Department of Energy, Office of High Energy Physics. 
Z.H. wants to thank the Fermilab LPC Guests and Visitors program. 
A.G., J.N., and J.D. are supported by the DOE Office of Science, Award No. DE-SC0023524, FermiForward Discovery Group, LLC under Contract No. 89243024CSC000002 with the U.S. Department of Energy, Office of Science, Office of High Energy Physics, LDRD L2024-066-1, Fermilab, DOE Office of Science, Office of High Energy Physics ``Designing efficient edge AI with physics phenomena'' Project (DE-FOA-0002705), DOE Office of Science, Office of Advanced Scientific Computing Research under the ``Real-time Data Reduction Codesign at the Extreme Edge for Science'' Project (DE-FOA-0002501).
J.D. is also supported by the Research Corporation for Science Advancement (RCSA) under grant \#CS-CSA-2023-109, Alfred P. Sloan Foundation under grant \#FG-2023-20452, U.S. Department of Energy (DOE), Office of Science, Office of High Energy Physics Early Career Research program under Award No. DE-SC0021187, and the U.S. National Science Foundation (NSF) Harnessing the Data Revolution (HDR) Institute for Accelerating AI Algorithms for Data Driven Discovery (A3D3) under Cooperative Agreement PHY-2117997.
R.K. and J.N. are also supported by the AI2050 program at Schmidt Futures (Grant G-23-64934).
This work was performed using the National Research Platform Nautilus HyperCluster supported by NSF awards CNS-1730158, ACI-1540112, ACI-1541349, OAC-1826967, OAC-2112167, CNS-2100237, CNS-2120019, the University of California Office of the President, and the University of California San Diego's California Institute for Telecommunications and Information Technology/Qualcomm Institute.
\end{ack}

\bibliographystyle{cms_unsrt}
\bibliography{bibliography.bib}

@unpublished{HEP-JEPA,
    author = "Bardhan, Jai and Agrawal, Radhikesh and Tilak, Abhiram and Neeraj, Cyrin and Mitra, Subhadip",
    title = "{HEP-JEPA: A foundation model for collider physics using joint embedding predictive architecture}",
    eprint = "2502.03933",
    archivePrefix = "arXiv",
    primaryClass = "cs.LG",
    month = "2",
    year = "2025"
}

@article{Mikuni:2024qsr,
    author = "Mikuni, Vinicius and Nachman, Benjamin",
    title = "{Solving key challenges in collider physics with foundation models}",
    eprint = "2404.16091",
    archivePrefix = "arXiv",
    primaryClass = "hep-ph",
    doi = "10.1103/PhysRevD.111.L051504",
    journal = "Phys. Rev. D",
    volume = "111",
    pages = "L051504",
    year = "2025"
}

@article{Mikuni:2025tar,
    author = "Mikuni, Vinicius and Nachman, Benjamin",
    title = "{Method to simultaneously facilitate all jet physics tasks}",
    eprint = "2502.14652",
    archivePrefix = "arXiv",
    primaryClass = "hep-ph",
    doi = "10.1103/PhysRevD.111.054015",
    journal = "Phys. Rev. D",
    volume = "111",
    number = "5",
    pages = "054015",
    year = "2025"
}

@unpublished{Li:2024htp,
    author = "Li, Congqiao and others",
    title = "{Accelerating Resonance Searches via Signature-Oriented Pre-training}",
    eprint = "2405.12972",
    archivePrefix = "arXiv",
    primaryClass = "hep-ph",
    reportNumber = "FERMILAB-PUB-24-0699-V",
    year = "2024"
}

@inproceedings{Zhao:2024kry,
    author = "Zhao, Zihan and Mokhtar, Farouk and Kansal, Raghav and Li, Haoyang and Duarte, Javier",
    title = "{Large-Scale Pretraining and Finetuning for Efficient Jet Classification in Particle Physics}",
    booktitle = "{22nd International Workshop on Advanced Computing and Analysis Techniques in Physics Research}",
    eprint = "2408.09343",
    archivePrefix = "arXiv",
    primaryClass = "hep-ex",
    year = "2024"
}

@article{Mokhtar:2025zqs,
    author = "Mokhtar, Farouk and Pata, Joosep and Garcia, Dolores and Wulff, Eric and Zhang, Mengke and Kagan, Michael and Duarte, Javier",
    title = "{Fine-tuning machine-learned particle-flow reconstruction for new detector geometries in future colliders}",
    eprint = "2503.00131",
    archivePrefix = "arXiv",
    primaryClass = "hep-ex",
    doi = "10.1103/PhysRevD.111.092015",
    journal = "Phys. Rev. D",
    volume = "111",
    number = "9",
    pages = "092015",
    year = "2025"
}

@article{Resimulation,
    author = "Harris, Philip and Krupa, Jeffrey and Kagan, Michael and Maier, Benedikt and Woodward, Nathaniel",
    title = "{Resimulation-based self-supervised learning for pretraining physics foundation models}",
    eprint = "2403.07066",
    archivePrefix = "arXiv",
    primaryClass = "hep-ph",
    doi = "10.1103/PhysRevD.111.032010",
    journal = "Phys. Rev. D",
    volume = "111",
    pages = "032010",
    year = "2025"
}

@inproceedings{DINO,
  author       = {Mathilde Caron and
                  Hugo Touvron and
                  Ishan Misra and
                  Herv{\'{e}} J{\'{e}}gou and
                  Julien Mairal and
                  Piotr Bojanowski and
                  Armand Joulin},
  title        = {Emerging Properties in Self-Supervised Vision Transformers},
  volume       = {abs/2104.14294},
  year         = {2021},
  eprinttype    = {arXiv},
  eprint       = {2104.14294},
  booktitle={2021 IEEE/CVF International Conference on Computer Vision (ICCV)},
  pages={9630},
  doi={10.1109/ICCV48922.2021.00951}
}

@article{DINOv2,
      title={{DINOv2}: Learning Robust Visual Features without Supervision}, 
      author={Maxime Oquab and Timoth{\'e}e Darcet and Théo Moutakanni and Huy Vo and Marc Szafraniec and Vasil Khalidov and Pierre Fernandez and Daniel Haziza and Francisco Massa and Alaaeldin El-Nouby and Mahmoud Assran and Nicolas Ballas and Wojciech Galuba and Russell Howes and Po-Yao Huang and Shang-Wen Li and Ishan Misra and Michael Rabbat and Vasu Sharma and Gabriel Synnaeve and Hu Xu and Herv{\'e} Jegou and Julien Mairal and Patrick Labatut and Armand Joulin and Piotr Bojanowski},
      year={2024},
      eprint={2304.07193},
      archivePrefix={arXiv},
      primaryClass={cs.CV},
      journal={Trans. Mach. Learn. Res.},
      url={https://openreview.net/forum?id=a68SUt6zFt},
}

@article{kt-clustering,
    title = {Longitudinally-invariant \kt-clustering algorithms for hadron-hadron collisions},
    reportNumber = "CERN-TH-6775-93, LU-TP-93-2",
    journal = "Nucl. Phys. B",
    volume = {406},
    pages = {187},
    year = {1993},
    issn = {0550-3213},
    doi = {10.1016/0550-3213(93)90166-M},
    author = {S. Catani and Yu.L. Dokshitzer and M.H. Seymour and B.R. Webber},
}

@article{FastJet,
    author = "Cacciari, Matteo and Salam, Gavin P. and Soyez, Gregory",
    title = "{FastJet User Manual}",
    eprint = "1111.6097",
    archivePrefix = "arXiv",
    primaryClass = "hep-ph",
    reportNumber = "CERN-PH-TH-2011-297",
    doi = "10.1140/epjc/s10052-012-1896-2",
    journal = "Eur. Phys. J. C",
    volume = "72",
    pages = "1896",
    year = "2012"
}

@inproceedings{PyTorch,
  author       = {Adam Paszke and
                  Sam Gross and
                  Francisco Massa and
                  Adam Lerer and
                  James Bradbury and
                  Gregory Chanan and
                  Trevor Killeen and
                  Zeming Lin and
                  Natalia Gimelshein and
                  Luca Antiga and
                  Alban Desmaison and
                  Andreas K{\"{o}}pf and
                  Edward Z. Yang and
                  Zach DeVito and
                  Martin Raison and
                  Alykhan Tejani and
                  Sasank Chilamkurthy and
                  Benoit Steiner and
                  Lu Fang and
                  Junjie Bai and
                  Soumith Chintala},
  title        = {{PyTorch}: An Imperative Style, High-Performance Deep Learning Library},
  year         = {2019},
  eprinttype    = {arXiv},
  eprint       = {1912.01703},
	booktitle = {Advances in Neural Information Processing Systems},
	editor = {H. Wallach and H. Larochelle and A. Beygelzimer and F. d\textquotesingle Alch\'{e}-Buc and E. Fox and R. Garnett},
	publisher = {Curran Associates, Inc.},
	url = {https://proceedings.neurips.cc/paper_files/paper/2019/file/bdbca288fee7f92f2bfa9f7012727740-Paper.pdf},
	volume = {32},
}

@inproceedings{ParT,
    author = "Qu, Huilin and Li, Congqiao and Qian, Sitian",
    title = "{Particle Transformer for Jet Tagging}",
    eprint = "2202.03772",
    archivePrefix = "arXiv",
    primaryClass = "hep-ph",
    year = "2022",
  pages = 	 {18281},
  editor = 	 {Chaudhuri, Kamalika and Jegelka, Stefanie and Song, Le and Szepesvari, Csaba and Niu, Gang and Sabato, Sivan},
  volume = 	 {162},
  booktitle = 	 {Proceedings of the 39th International Conference on Machine Learning},
  url = 	 {https://proceedings.mlr.press/v162/qu22b.html},
}

@misc{JetClass,
  author       = {Qu, Huilin and
                  Li, Congqiao and
                  Qian, Sitian},
  title        = {{JetClass}: A Large-Scale Dataset for Deep Learning
                   in Jet Physics
                  },
  year         = 2022,
  publisher    = {Zenodo},
  version      = {1.0.0},
  doi          = {10.5281/zenodo.6619768},
}

@inproceedings{JetNet-dataset,
  author = {Kansal, Raghav and Duarte, Javier and Su, Hao and Orzari, Breno and Tomei, Thiago and Pierini, Maurizio and Touranakou, Mary and Vlimant, Jean-Roch and Gunopulos, Dimitrios},
  booktitle = "{Advances in Neural Information Processing Systems}",
  editor = {M. Ranzato and A. Beygelzimer and Y. Dauphin and P.S. Liang and J. Wortman Vaughan},
  pages = {23858},
  publisher = {Curran Associates, Inc.},
  title = {Particle Cloud Generation with Message Passing Generative Adversarial Networks},
  url = {https://proceedings.neurips.cc/paper_files/paper/2021/file/c8512d142a2d849725f31a9a7a361ab9-Paper.pdf},
  volume = {34},
  year = {2021},
  eprint = {2106.11535},
  archivePrefix = {arXiv},
}

@article{LorentzNet,
   title={An efficient Lorentz equivariant graph neural network for jet tagging},
   volume={07},
   DOI={10.1007/jhep07(2022)030},
eprint = "2201.08187",
    archivePrefix = "arXiv",
    primaryClass = "hep-ph",
   journal={JHEP},
   author={Gong, Shiqi and Meng, Qi and Zhang, Jue and Qu, Huilin and Li, Congqiao and Qian, Sitian and Du, Weitao and Ma, Zhi-Ming and Liu, Tie-Yan},
   year={2022}
}

@inproceedings{interaction-networks,
      title={Interaction Networks for Learning about Objects, Relations and Physics}, 
      author={Peter W. Battaglia and Razvan Pascanu and Matthew Lai and Danilo Rezende and Koray Kavukcuoglu},
      year={2016},
      eprint={1612.00222},
      archivePrefix={arXiv},
      primaryClass={cs.AI},
	booktitle = {Advances in Neural Information Processing Systems},
	editor = {D. Lee and M. Sugiyama and U. Luxburg and I. Guyon and R. Garnett},
	publisher = {Curran Associates, Inc.},
	url = {https://proceedings.neurips.cc/paper_files/paper/2016/file/3147da8ab4a0437c15ef51a5cc7f2dc4-Paper.pdf},
	volume = {29},
}

@article{EFN,
    author = "Komiske, Patrick T. and Metodiev, Eric M. and Thaler, Jesse",
    title = "{Energy Flow Networks: Deep Sets for Particle Jets}",
    eprint = "1810.05165",
    archivePrefix = "arXiv",
    primaryClass = "hep-ph",
    reportNumber = "MIT-CTP 5064",
    doi = "10.1007/JHEP01(2019)121",
    journal = "JHEP",
    volume = "01",
    pages = "121",
    year = "2019"
}

@misc{BERT,
      title={BERT: Pre-training of Deep Bidirectional Transformers for Language Understanding}, 
      author={Jacob Devlin and Ming-Wei Chang and Kenton Lee and Kristina Toutanova},
      year={2019},
      eprint={1810.04805},
      archivePrefix={arXiv},
      primaryClass={cs.CL},
      url={https://arxiv.org/abs/1810.04805}, 
}

@inproceedings{SSL-single-image,
      title={A critical analysis of self-supervision, or what we can learn from a single image}, 
      author={Yuki M. Asano and Christian Rupprecht and Andrea Vedaldi},
      year={2020},
      eprint={1904.13132},
      archivePrefix={arXiv},
      primaryClass={cs.CV},
booktitle={International Conference on Learning Representations},
url={https://openreview.net/forum?id=B1esx6EYvr}
}

@inproceedings{SSL-robustness,
      title={Using Self-Supervised Learning Can Improve Model Robustness and Uncertainty}, 
      author={Dan Hendrycks and Mantas Mazeika and Saurav Kadavath and Dawn Song},
      year={2019},
      eprint={1906.12340},
      archivePrefix={arXiv},
      primaryClass={cs.LG},
	booktitle = {Advances in Neural Information Processing Systems},
	editor = {H. Wallach and H. Larochelle and A. Beygelzimer and F. d\textquotesingle Alch\'{e}-Buc and E. Fox and R. Garnett},
	publisher = {Curran Associates, Inc.},
	url = {https://proceedings.neurips.cc/paper_files/paper/2019/file/a2b15837edac15df90721968986f7f8e-Paper.pdf},
	volume = {32},
}

@article{SSL-survey,
      title={A Survey on Self-supervised Learning: Algorithms, Applications, and Future Trends}, 
      author={Jie Gui and Tuo Chen and Jing Zhang and Qiong Cao and Zhenan Sun and Hao Luo and Dacheng Tao},
      year={2024},
      eprint={2301.05712},
      archivePrefix={arXiv},
      primaryClass={cs.LG},
  journal={IEEE Trans. Pattern Anal. Mach. Intell.},
  volume={46},
  pages={9052},
  doi={10.1109/TPAMI.2024.3415112}
}

@inproceedings{SimCLR,
      title={A Simple Framework for Contrastive Learning of Visual Representations}, 
      author={Ting Chen and Simon Kornblith and Mohammad Norouzi and Geoffrey Hinton},
      year={2020},
      eprint={2002.05709},
      archivePrefix={arXiv},
      primaryClass={cs.LG},
  booktitle = 	 {Proceedings of the 37th International Conference on Machine Learning},
  pages = 	 {1597},
  editor = 	 {III, Hal Daumé and Singh, Aarti},
  volume = 	 {119},
  url = 	 {https://proceedings.mlr.press/v119/chen20j.html},
}

@inproceedings{MOCO,
      title={Momentum Contrast for Unsupervised Visual Representation Learning}, 
      author={Kaiming He and Haoqi Fan and Yuxin Wu and Saining Xie and Ross Girshick},
      year={2020},
      eprint={1911.05722},
      archivePrefix={arXiv},
      primaryClass={cs.CV},
      url={https://arxiv.org/abs/1911.05722},
  booktitle={2020 IEEE/CVF Conference on Computer Vision and Pattern Recognition (CVPR)}, 
  pages={9726},
  doi={10.1109/CVPR42600.2020.00975}
}

@inproceedings{SwAV,
      title={Unsupervised Learning of Visual Features by Contrasting Cluster Assignments}, 
      author={Mathilde Caron and Ishan Misra and Julien Mairal and Priya Goyal and Piotr Bojanowski and Armand Joulin},
      year={2020},
      eprint={2006.09882},
      archivePrefix={arXiv},
      primaryClass={cs.CV},
	booktitle = {Advances in Neural Information Processing Systems},
	editor = {H. Larochelle and M. Ranzato and R. Hadsell and M.F. Balcan and H. Lin},
	pages = {9912},
	publisher = {Curran Associates, Inc.},
	url = {https://proceedings.neurips.cc/paper_files/paper/2020/file/70feb62b69f16e0238f741fab228fec2-Paper.pdf},
	volume = {33},
}

@inproceedings{J-JEPA,
    author = "Katel, Subash and Li, Haoyang and Zhao, Zihan and Mokhtar, Farouk and Duarte, Javier and Kansal, Raghav",
    title = "{Learning Symmetry-Independent Jet Representations via Jet-Based Joint Embedding Predictive Architecture}",
    booktitle = "{Machine Learning and the Physical Sciences Workshop at the 38th Conference on Neural Information Processing Systems}",
    eprint = "2412.05333",
    archivePrefix = "arXiv",
    primaryClass = "hep-ph",
    year = "2024"
}

@unpublished{DINOv3,
      title={{DINOv3}}, 
      author={Oriane Sim{\'e}oni and Huy V. Vo and Maximilian Seitzer and Federico Baldassarre and Maxime Oquab and Cijo Jose and Vasil Khalidov and Marc Szafraniec and Seungeun Yi and Michaël Ramamonjisoa and Francisco Massa and Daniel Haziza and Luca Wehrstedt and Jianyuan Wang and Timoth{\'e}e Darcet and Th{\'e}o Moutakanni and Leonel Sentana and Claire Roberts and Andrea Vedaldi and Jamie Tolan and John Brandt and Camille Couprie and Julien Mairal and Herv{\'e} J{\'e}gou and Patrick Labatut and Piotr Bojanowski},
      year={2025},
      eprint={2508.10104},
      archivePrefix={arXiv},
      primaryClass={cs.CV},
}

@article{MPM,
    author = "Golling, Tobias and Heinrich, Lukas and Kagan, Michael and Klein, Samuel and Leigh, Matthew and Osadchy, Margarita and Raine, John Andrew",
    title = "{Masked particle modeling on sets: towards self-supervised high energy physics foundation models}",
    eprint = "2401.13537",
    archivePrefix = "arXiv",
    primaryClass = "hep-ph",
    doi = "10.1088/2632-2153/ad64a8",
    journal = "Mach. Learn. Sci. Tech.",
    volume = "5",
    pages = "035074",
    year = "2024"
}

@article{MPM-improved,
    author = "Leigh, Matthew and Klein, Samuel and Charton, Fran{\c{c}}ois and Golling, Tobias and Heinrich, Lukas and Kagan, Michael and Ochoa, In{\^e}s and Osadchy, Margarita",
    eprint = "2409.12589",
    archivePrefix = "arXiv",
    primaryClass = "hep-ph",
    year = "2025",
	doi = {10.1088/2632-2153/addb98},
	journal = {Mach. Learn. Sci. Tech.},
	pages = {025075},
	title = {Is tokenization needed for masked particle modeling?},
	volume = {6},
}

@article{JetCLR,
    author = "Dillon, Barry M. and Kasieczka, Gregor and Olischlager, Hans and Plehn, Tilman and Sorrenson, Peter and Vogel, Lorenz",
    title = "{Symmetries, safety, and self-supervision}",
    eprint = "2108.04253",
    archivePrefix = "arXiv",
    primaryClass = "hep-ph",
    doi = "10.21468/SciPostPhys.12.6.188",
    journal = "SciPost Phys.",
    volume = "12",
    pages = "188",
    year = "2022"
}

@article{OmniJet-alpha,
   title={{OmniJet-$\alpha$}: the first cross-task foundation model for particle physics},
   volume={5},
   DOI={10.1088/2632-2153/ad66ad},
   journal={Mach. Learn. Sci. Tech.},
   author={Birk, Joschka and Hallin, Anna and Kasieczka, Gregor},
   year={2024},
pages={035031},
    eprint = "2403.05618",
    archivePrefix = "arXiv",
    primaryClass = "hep-ph",
}

@misc{ViT,
      title={An Image is Worth 16x16 Words: Transformers for Image Recognition at Scale}, 
      author={Alexey Dosovitskiy and Lucas Beyer and Alexander Kolesnikov and Dirk Weissenborn and Xiaohua Zhai and Thomas Unterthiner and Mostafa Dehghani and Matthias Minderer and Georg Heigold and Sylvain Gelly and Jakob Uszkoreit and Neil Houlsby},
      year={2021},
      eprint={2010.11929},
      archivePrefix={arXiv},
      primaryClass={cs.CV},
      url={https://arxiv.org/abs/2010.11929}, 
}

@inproceedings{transformer,
      eprint={1706.03762},
      archivePrefix={arXiv},
      primaryClass={cs.CL},
	author = {Vaswani, Ashish and Shazeer, Noam and Parmar, Niki and Uszkoreit, Jakob and Jones, Llion and Gomez, Aidan N and Kaiser, \L ukasz and Polosukhin, Illia},
	booktitle = {Advances in Neural Information Processing Systems},
	editor = {I. Guyon and U. Von Luxburg and S. Bengio and H. Wallach and R. Fergus and S. Vishwanathan and R. Garnett},
	publisher = {Curran Associates, Inc.},
	title = {Attention is All you Need},
	url = {https://proceedings.neurips.cc/paper_files/paper/2017/file/3f5ee243547dee91fbd053c1c4a845aa-Paper.pdf},
	volume = {30},
	year = {2017},
}

@inbook{jets,
author = {Simone Marzani},
title = {Jets at Colliders},
booktitle = {Instrumentation and Techniques in High Energy Physics},
chapter = {Chapter 6},
pages = {179-211},
doi = {10.1142/9789819801107_0006},
URL = {https://www.worldscientific.com/doi/abs/10.1142/9789819801107_0006},
eprint = {https://www.worldscientific.com/doi/pdf/10.1142/9789819801107_0006}
}

@inproceedings{KoLeo,
      title={Spreading vectors for similarity search}, 
      author={Alexandre Sablayrolles and Matthijs Douze and Cordelia Schmid and Herv{\'e} J{\'e}gou},
      year={2019},
      eprint={1806.03198},
      archivePrefix={arXiv},
      primaryClass={stat.ML},
booktitle={International Conference on Learning Representations},
url={https://openreview.net/forum?id=SkGuG2R5tm},
}

@inproceedings{SK-centering,
 author = {Caron, Mathilde and Misra, Ishan and Mairal, Julien and Goyal, Priya and Bojanowski, Piotr and Joulin, Armand},
 booktitle = {Advances in Neural Information Processing Systems},
 editor = {H. Larochelle and M. Ranzato and R. Hadsell and M.F. Balcan and H. Lin},
 pages = {9912},
 publisher = {Curran Associates, Inc.},
 title = {Unsupervised Learning of Visual Features by Contrasting Cluster Assignments},
 url = {https://proceedings.neurips.cc/paper_files/paper/2020/file/70feb62b69f16e0238f741fab228fec2-Paper.pdf},
 volume = {33},
 year = {2020}
}

@article{k-NN,
 author = {Evelyn Fix and J. L. Hodges},
 journal = {Int. Stat. Rev.},
 pages = {238},
 title = {Discriminatory Analysis. Nonparametric Discrimination: Consistency Properties},
doi = {10.2307/1403797},
 volume = {57},
 year = {1989}
}

@article{BDT,
 ISSN = {00905364, 21688966},
 URL = {http://www.jstor.org/stable/2699986},
 author = {Jerome H. Friedman},
 journal = {Ann. Stat.},
 pages = {1189},
 publisher = {Institute of Mathematical Statistics},
 title = {Greedy Function Approximation: A Gradient Boosting Machine},
 volume = {29},
 year = {2001}
}

@article{t-SNE,
  author  = {Laurens van der Maaten and Geoffrey Hinton},
  title   = {Visualizing Data using {t-SNE}},
  journal = {J. Mach. Learn. Res.},
  year    = {2008},
  volume  = {9},
  pages   = {2579},
  url     = {http://jmlr.org/papers/v9/vandermaaten08a.html}
}

@misc{AdamW,
      title={Decoupled Weight Decay Regularization}, 
      author={Ilya Loshchilov and Frank Hutter},
      year={2019},
      eprint={1711.05101},
      archivePrefix={arXiv},
      primaryClass={cs.LG},
booktitle={International Conference on Learning Representations},
url={https://openreview.net/forum?id=Bkg6RiCqY7},
}

@Misc{accelerate,
  title =        {Accelerate: Training and inference at scale made simple, efficient and adaptable.},
  author =       {Sylvain Gugger and Lysandre Debut and Thomas Wolf and Philipp Schmid and Zachary Mueller and Sourab Mangrulkar and Marc Sun and Benjamin Bossan},
  howpublished = {\url{https://github.com/huggingface/accelerate}},
  year =         {2022}
}

@article{PELICAN,
    author = "Bogatskiy, Alexander and Hoffman, Timothy and Miller, David W. and Offermann, Jan T. and Liu, Xiaoyang",
    title = "{Explainable equivariant neural networks for particle physics: PELICAN}",
    eprint = "2307.16506",
    archivePrefix = "arXiv",
    primaryClass = "hep-ph",
    doi = "10.1007/JHEP03(2024)113",
    journal = "JHEP",
    volume = "03",
    pages = "113",
    year = "2024"
}

@inproceedings{L-GATr,
    author = "Spinner, Jonas and Bres{\'o}, Victor and de Haan, Pim and Plehn, Tilman and Thaler, Jesse and Brehmer, Johann",
    title = "{Lorentz-Equivariant Geometric Algebra Transformers for High-Energy Physics}",
    eprint = "2405.14806",
    archivePrefix = "arXiv",
    primaryClass = "physics.data-an",
    reportNumber = "MIT-CTP/5723",
    year = "2024",
	booktitle = {Advances in Neural Information Processing Systems},
	editor = {A. Globerson and L. Mackey and D. Belgrave and A. Fan and U. Paquet and J. Tomczak and C. Zhang},
	pages = {22178},
	publisher = {Curran Associates, Inc.},
	url = {https://proceedings.neurips.cc/paper_files/paper/2024/file/277628cff838927d869cd1f671328ce0-Paper-Conference.pdf},
	volume = {37},
}

@misc{hepmllivingreview,
    Author = "{HEP ML Community}",
    title = "{A Living Review of Machine Learning for Particle Physics}",
    url={https://iml-wg.github.io/HEPML-LivingReview/}
}

@unpublished{Lorentz-equivariant-transformer-LHC,
      title={A {Lorentz}-Equivariant Transformer for All of the {LHC}}, 
      author={Johann Brehmer and Víctor Bresó and Pim de Haan and Tilman Plehn and Huilin Qu and Jonas Spinner and Jesse Thaler},
      year={2024},
      eprint={2411.00446},
      archivePrefix={arXiv},
      primaryClass={hep-ph},
note = "Submitted to \emph{SciPost Phys.}"
}

@article{scikit-learn,
  title={Scikit-learn: Machine Learning in {P}ython},
  author={Pedregosa, F. and Varoquaux, G. and Gramfort, A. and Michel, V.
          and Thirion, B. and Grisel, O. and Blondel, M. and Prettenhofer, P.
          and Weiss, R. and Dubourg, V. and Vanderplas, J. and Passos, A. and
          Cournapeau, D. and Brucher, M. and Perrot, M. and Duchesnay, E.},
  journal={J. Mach. Learn. Res.},
  volume={12},
  pages={2825},
  year={2011},
  url     = {http://jmlr.org/papers/v12/pedregosa11a.html}
}

@Article{matplotlib,
  Author    = {Hunter, J. D.},
  Title     = {Matplotlib: A {2D} graphics environment},
  Journal   = {Comput. Sci. Eng.},
  Volume    = {9},
  Pages     = {90},
  doi       = {10.1109/MCSE.2007.55},
  year      = 2007
}

@Article{Numpy,
 title         = {Array programming with {NumPy}},
 author        = {Charles R. Harris and K. Jarrod Millman and St{\'{e}}fan J.
                 van der Walt and Ralf Gommers and Pauli Virtanen and David
                 Cournapeau and Eric Wieser and Julian Taylor and Sebastian
                 Berg and Nathaniel J. Smith and Robert Kern and Matti Picus
                 and Stephan Hoyer and Marten H. van Kerkwijk and Matthew
                 Brett and Allan Haldane and Jaime Fern{\'{a}}ndez del
                 R{\'{i}}o and Mark Wiebe and Pearu Peterson and Pierre
                 G{\'{e}}rard-Marchant and Kevin Sheppard and Tyler Reddy and
                 Warren Weckesser and Hameer Abbasi and Christoph Gohlke and
                 Travis E. Oliphant},
 year          = {2020},
 journal       = {Nature},
 volume        = {585},
 number        = {7825},
 pages         = {357},
 doi           = {10.1038/s41586-020-2649-2},
eprint={2006.10256}
}

@inproceedings{NRP,
author = {Weitzel, Derek and Graves, Ashton and Albin, Sam and Zhu, Huijun and Wuerthwein, Frank and Tatineni, Mahidhar and Mishin, Dmitry and Khoda, Elham and Sada, Mohammad and Smarr, Larry and DeFanti, Thomas and Graham, John},
title = {The {National Research Platform}: Stretched, Multi-Tenant, Scientific {Kubernetes} Cluster},
year = {2025},
isbn = {9798400713989},
publisher = {Association for Computing Machinery},
doi = {10.1145/3708035.3736060},
booktitle = {Practice and Experience in Advanced Research Computing 2025: The Power of Collaboration},
}

@article{Andersson:1987pr,
    author = "Andersson, Bo",
    title = "{The Lund Model}",
    doi = "10.1016/0375-9474(87)90510-0",
    journal = "Nucl. Phys. A",
    volume = "461",
    pages = "513C",
    year = "1987"
}

@article{PhysRevD.105.054012,
    author = "Nagy, Zoltan and Soper, Davison E.",
    title = "{Multivariable evolution in final state parton shower algorithms}",
    eprint = "2201.08056",
    archivePrefix = "arXiv",
    primaryClass = "hep-ph",
    reportNumber = "DESY-22-009",
    doi = "10.1103/PhysRevD.105.054012",
    journal = "Phys. Rev. D",
    volume = "105",
    pages = "054012",
    year = "2022"
}

@unpublished{CMS:2025eyd,
    author = "Hayrapetyan, Aram and others",
    collaboration = "CMS",
    title = "{A method for correcting the substructure of multiprong jets using the Lund jet plane}",
    eprint = "2507.07775",
    archivePrefix = "arXiv",
    primaryClass = "hep-ex",
    reportNumber = "CMS-JME-23-001, CERN-EP-2025-128",
    year = "2025",
note = "Submitted to \emph{JHEP}"
}

@article{Thaler:2010tr,
    author = "Thaler, Jesse and Van Tilburg, Ken",
    title = "{Identifying Boosted Objects with N-subjettiness}",
    eprint = "1011.2268",
    archivePrefix = "arXiv",
    primaryClass = "hep-ph",
    reportNumber = "MIT-CTP-4191",
    doi = "10.1007/JHEP03(2011)015",
    journal = "JHEP",
    volume = "03",
    pages = "015",
    year = "2011"
}

@article{EMA,
title = {Forecasting seasonals and trends by exponentially weighted moving averages},
journal = {Int. J. Forecast.},
volume = {20},
pages = {5},
year = {2004},
doi = {10.1016/j.ijforecast.2003.09.015},
author = {Charles C. Holt},
}

@inproceedings{LP-FT,
      title={Fine-Tuning can Distort Pretrained Features and Underperform Out-of-Distribution}, 
      author={Ananya Kumar and Aditi Raghunathan and Robbie Jones and Tengyu Ma and Percy Liang},
      year={2022},
      eprint={2202.10054},
      archivePrefix={arXiv},
      primaryClass={cs.LG},
booktitle={International Conference on Learning Representations},
url={https://openreview.net/forum?id=UYneFzXSJWh}
}

@inproceedings{ren2023preparetaskheadfinetuning,
      title={How to prepare your task head for finetuning}, 
      author={Yi Ren and Shangmin Guo and Wonho Bae and Danica J. Sutherland},
      year={2023},
      eprint={2302.05779},
      archivePrefix={arXiv},
      primaryClass={cs.LG},
  url={https://openreview.net/forum?id=gVOXZproe-e},
  booktitle={International Conference on Learning Representations},
}

@inproceedings{alain2018understandingintermediatelayersusing,
      title={Understanding intermediate layers using linear classifier probes}, 
      author={Guillaume Alain and Yoshua Bengio},
      year={2017},
      eprint={1610.01644},
      archivePrefix={arXiv},
      primaryClass={stat.ML},
booktitle = "International Conference on Learning Representations",
url={https://openreview.net/forum?id=ryF7rTqgl}
}

\medskip


\appendix

\section{Technical Appendices and Supplementary Material}
The code is provided in \url{https://github.com/zichunhao/RINO}.

\subsection{Model Architectures} \label{appendix:model-details}

\begin{figure}
    \centering
    \includegraphics[width=0.5\linewidth]{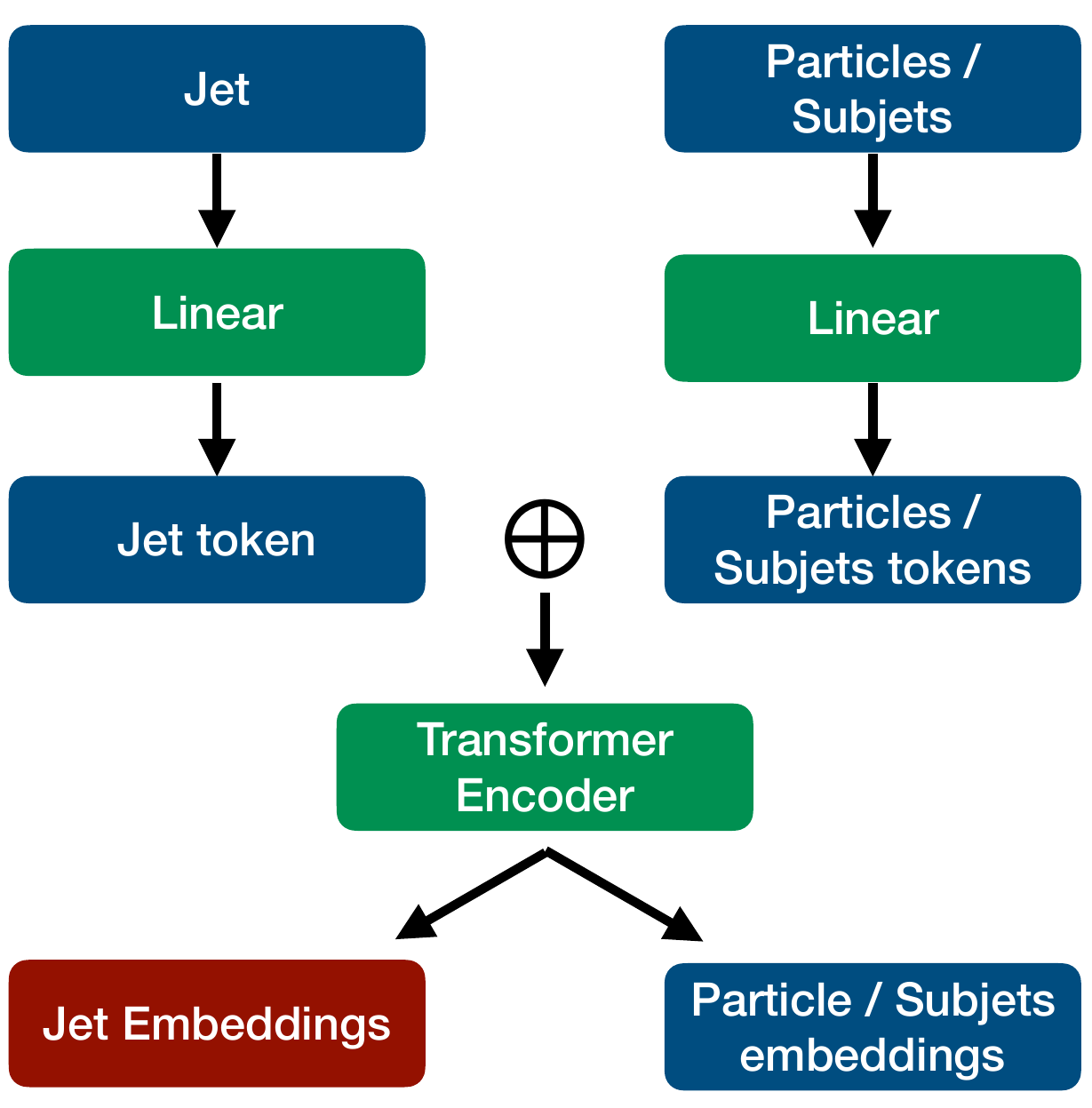}
    \caption{Model architecture of the transformer encoder backbone. The jet's representation is taken as the transformer embedding corresponding to the jet token.}
    \label{fig:model}
\end{figure}

\paragraph{Transformer Backbones}
All models employ transformer encoder architecture with GELU activation functions, pre-layer normalization, and jet-initialized class tokens that integrate jet kinematics for class token initialization with particle kinematics preprocessed following the particle transformer methodology~\cite{ParT}. 
Inspired by BERT~\cite{BERT} and ViT~\cite{ViT}, we take the embedding that corresponds to the jet token as the jet embedding, as shown in Figure~\ref{fig:model}.
The inputs are normalized manually for training stability. 
The transformer encoder architecture is implemented in four configurations with systematic scaling across embedding dimensions, attention heads, layers, and feedforward dimensions:
\begin{itemize}
    \item \textbf{Nano}: 32-dimensional encoder embeddings, 4 attention heads, 4 layers, 128-dimensional feedforward network (51,424 parameters)
    \item \textbf{Lite}: 64-dimensional encoder embeddings, 4 attention heads, 4 layers, 256-dimensional feedforward network (201,152 parameters)
    \item \textbf{Mini}: 128-dimensional encoder embeddings, 8 attention heads, 6 layers, 512-dimensional feedforward network (1,192,064 parameters)
    \item \textbf{Base}: 256-dimensional encoder embeddings, 8 attention heads, 6 layers, 1024-dimensional feedforward network (4,743,424 parameters)
\end{itemize}

\paragraph{DINO Projection Heads}
The DINO projection heads utilize multi-layer architectures with GELU activation and model-specific configurations:
\begin{itemize}
    \item \textbf{Nano}: 128-dimensional hidden layer producing 16-dimensional representations (6,272 parameters)
    \item \textbf{Lite}: 256-dimensional hidden layer producing 32-dimensional representations (24,832 parameters)
    \item \textbf{Mini}: 512-dimensional hidden layer producing 64-dimensional representations (98,816 parameters)
    \item \textbf{Base}: 1024-dimensional hidden layer producing 128-dimensional representations (394,240 parameters)
\end{itemize}
All projection heads incorporate $L_2$ normalization prior to the final projection layer and apply weight normalization to the output layer for training stabilization and enhanced representation quality.

\paragraph{Classification Heads}
For the RINO-Linear approach, we attach a linear head to the backbone. The numbers of parameters are 16, 32, 64, 128, and 256 parameters for nano, lite, mini, and base models respectively. For the RINO-MLP approach and supervised baselines, we employ two-layer multilayer perceptrons with GELU activation and dropout regularization. The architectures scale proportionally with model size:
\begin{itemize}
    \item \textbf{Nano}: First hidden layer of 8 dimensions, second hidden layer of 4 dimensions, dropout rate of 0.1 (304 parameters)
    \item \textbf{Lite}: First hidden layer of 16 dimensions, second hidden layer of 8 dimensions, dropout rate of 0.1 (1,184 parameters)
    \item \textbf{Mini}: First hidden layer of 32 dimensions, second hidden layer of 16 dimensions, dropout rate of 0.1 (4,672 parameters)
    \item \textbf{Base}: First hidden layer of 64 dimensions, second hidden layer of 32 dimensions, dropout rate of 0.1 (18,560 parameters)
\end{itemize}

\subsection{Training Details} \label{appendix:training-details}
\paragraph{RINO pretraining}
Global views use $\{1, 2, 3, 4\}$-cluster configurations, while student local views use $\{8, 16, 32, 64\}$-cluster configurations plus the original particle-level view. 
Teacher momentum starts at $0.992$ with cosine annealing to $1.0$. We use a teacher temperature of $0.07$ with cosine warmup from $0.04$ over $20$ epochs, a fixed student temperature of $0.10$. The Sinkhorn-Knopp algorithm~\cite{SK-centering} is used for centering, and the KoLeo regularization~\cite{KoLeo} is added with a weight of $1.0$ to prevent mode collapsing. 
Training employs the AdamW optimizer~\cite{AdamW} with a learning rate of $1 \times 10^{-4}$ and weight decay of $0.01$, following cosine annealing over 100 epochs. The base model uses 4 A100 GPUs, whereas the other four (nano, lite, and mini) models use 5 A10/3090 GPUs at the National Research Platform~\cite{NRP}. Multi-GPU training is achieved by \textsc{Huggingface Accelerate}~\cite{accelerate}. 
Each model takes 2-7 days to pretrain on the QCD jets from \jetclass.

\paragraph{RINO Fine-Tuning}
A task-specific classification head (linear layer for RINO-Linear or MLP for RINO-MLP) is attached to the pretrained backbone. Training employs the AdamW optimizer with weight decay of $0.01$ and an initial head learning rate of $1 \times 10^{-3}$. 
The backbone is frozen for the first 40 epochs, then unfrozen with a reduced learning rate of $1 \times 10^{-5}$ for the nano and lite models and $1 \times 10^{-6}$ for the mini and base models. 
The learning rate schedule follows a two-stage cosine strategy: 10-epoch warmup with learning factors scaling from $1 \times 10^{-4}$ to $1.0$, followed by 90-epoch cosine annealing from $1.0$ to $1 \times 10^{-3}$.
Binary cross-entropy loss is used with positive class weighting of 2.0 to address class imbalance. RINO-Linear models require 2 A10/3090 GPUs while RINO-MLP models require 4 A10/3090 GPUs, with training completing in under 60 minutes per model.

\paragraph{Supervised Baselines}
Training is performed from scratch using the AdamW optimizer with a learning rate of $1 \times 10^{-4}$ and weight decay of $0.01$, following cosine annealing for up to 1000 epochs with early stopping patience of 30 epochs.
All models use 2 A10/3090 GPUs. Each model takes less than 90 minutes to train. 

\subsection{Experiments} \label{appendix:experiments}
\begin{figure}[ht]
   \centering
   
   \includegraphics[width=0.29\linewidth]{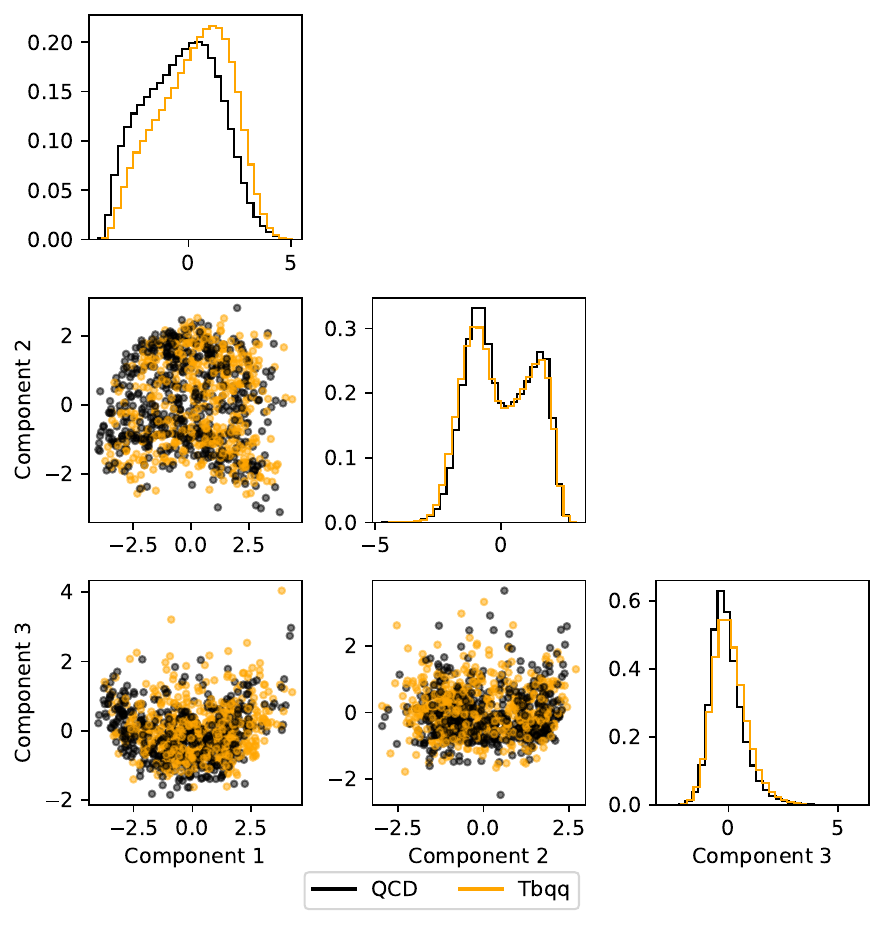}
   \includegraphics[width=0.4\linewidth]{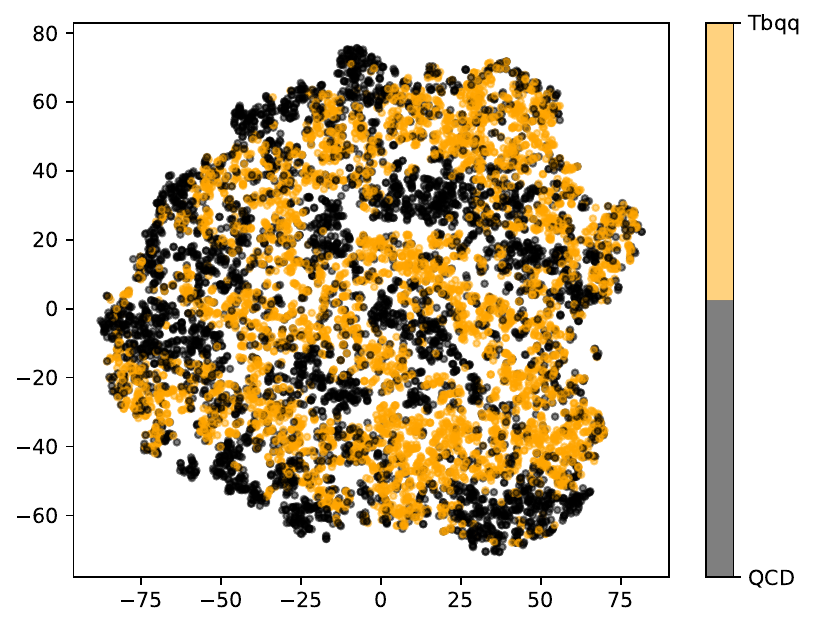}
   
   \includegraphics[width=0.29\linewidth]{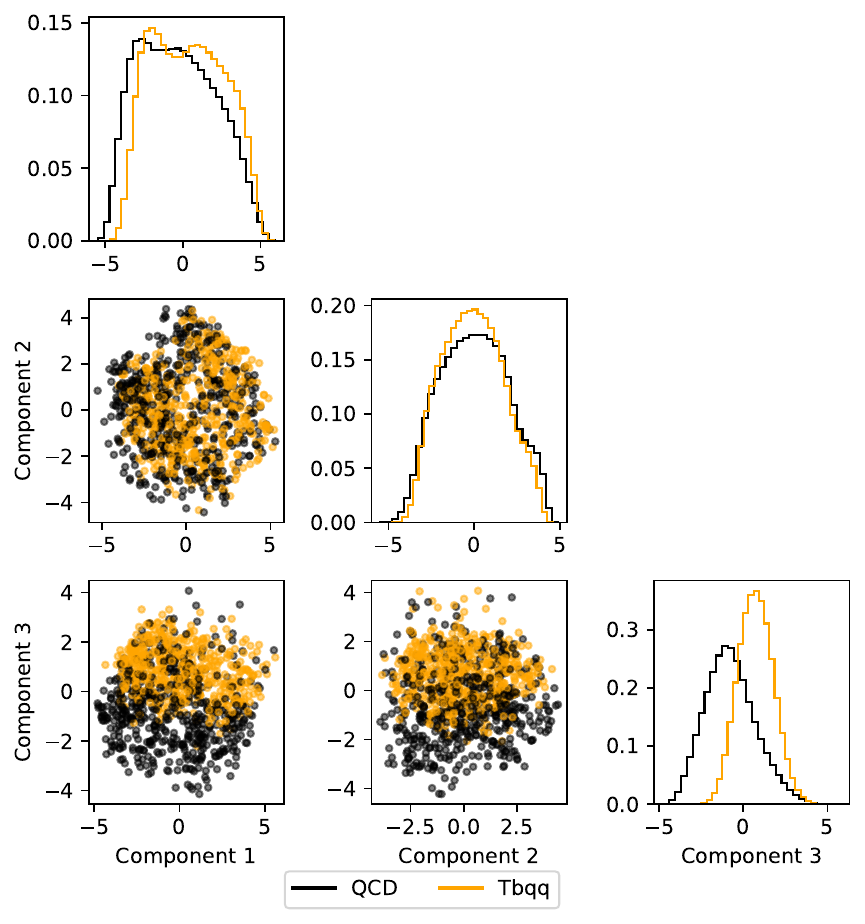}
   \includegraphics[width=0.4\linewidth]{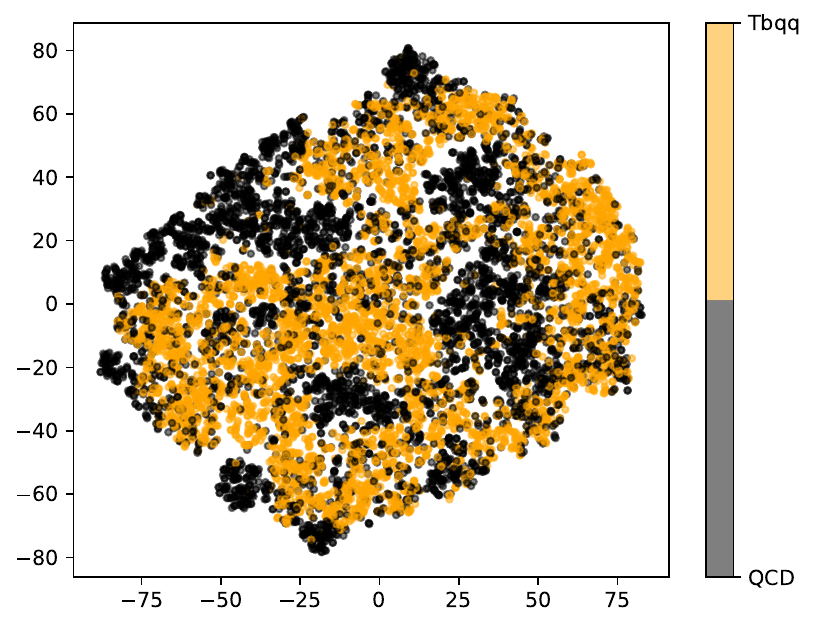}
   
   \includegraphics[width=0.29\linewidth]{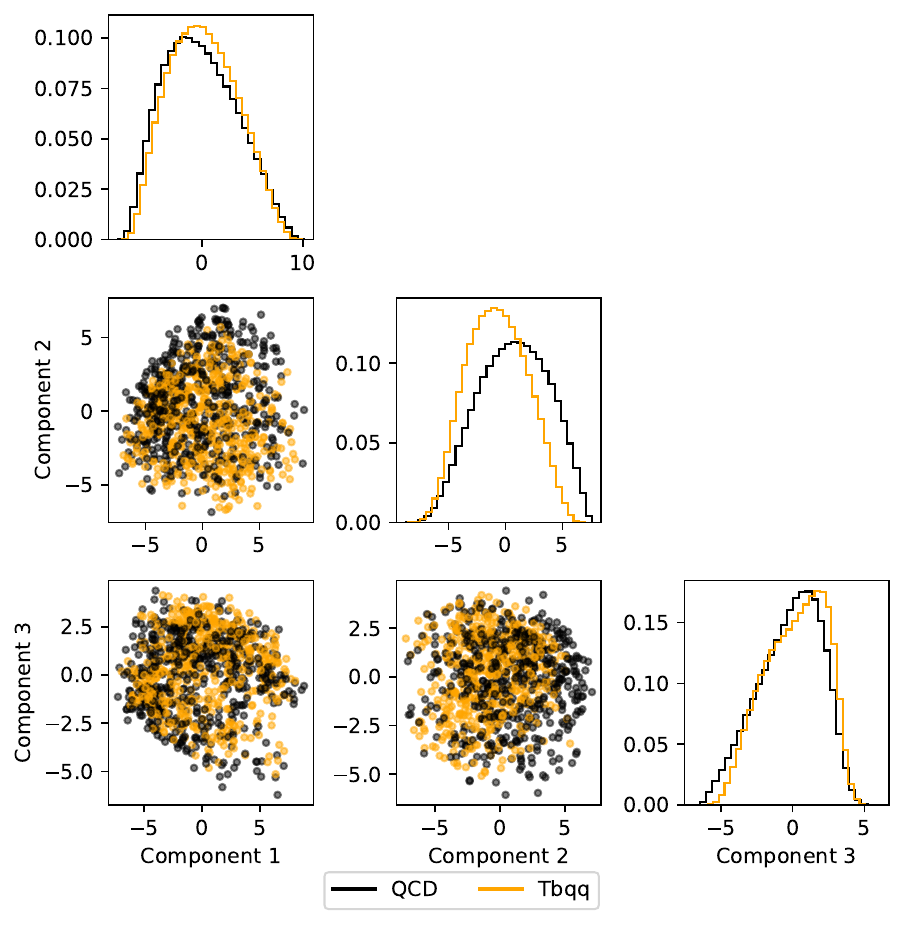}
   \includegraphics[width=0.4\linewidth]{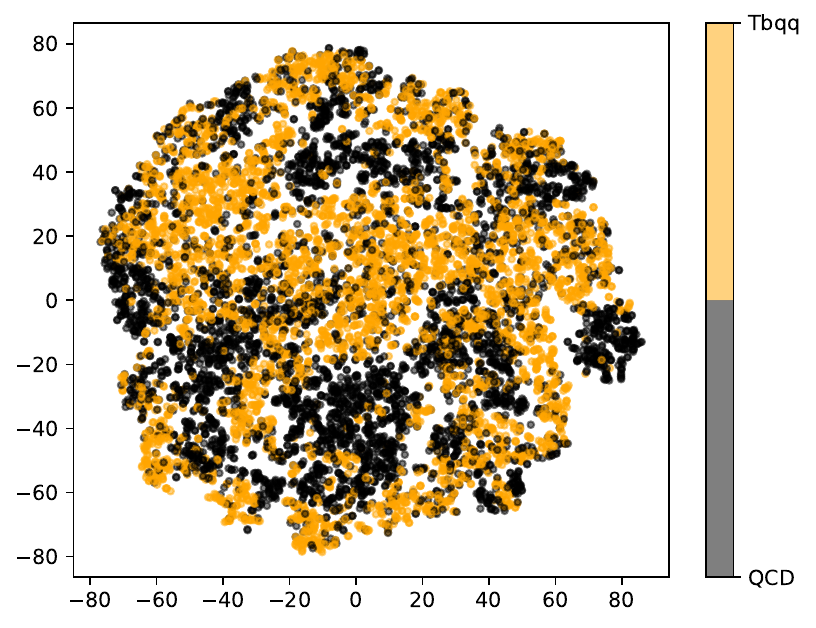}

   \includegraphics[width=0.29\linewidth]{figures/results/embeddings/base-cluster/pca_corner_plot.pdf}
   \includegraphics[width=0.4\linewidth]{figures/results/embeddings/base-cluster/tsne.pdf}
   
   \caption{
   Visualization of learned jet representations from the nano (row 1), lite (row 2), mini (row 3), and base (row 4) models  using PCA (left) and t-SNE embedding (right).}
   \label{fig:embedding-PCA-tSNE-all}
\end{figure}

\begin{figure}
    \centering
    \includegraphics[width=0.48\linewidth]{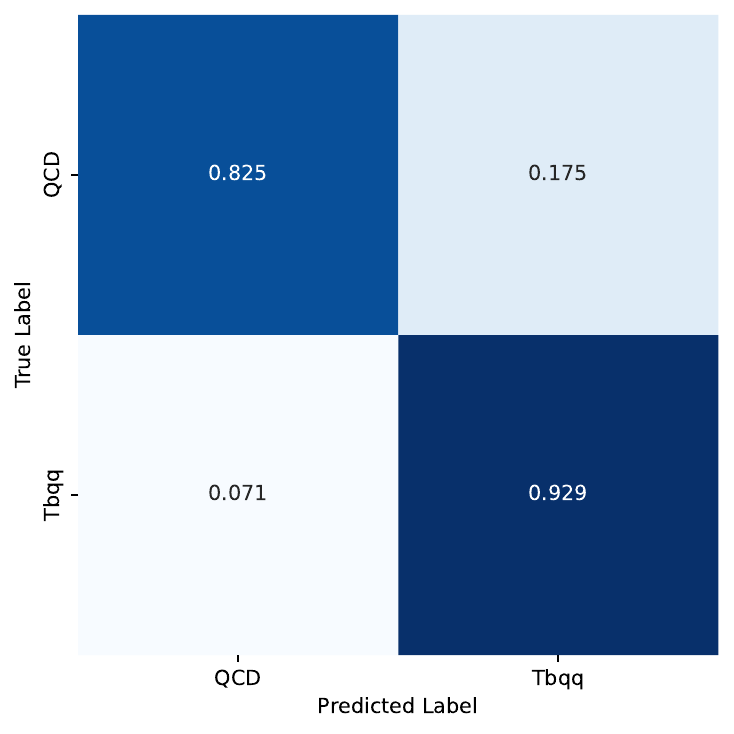}
    \includegraphics[width=0.48\linewidth]{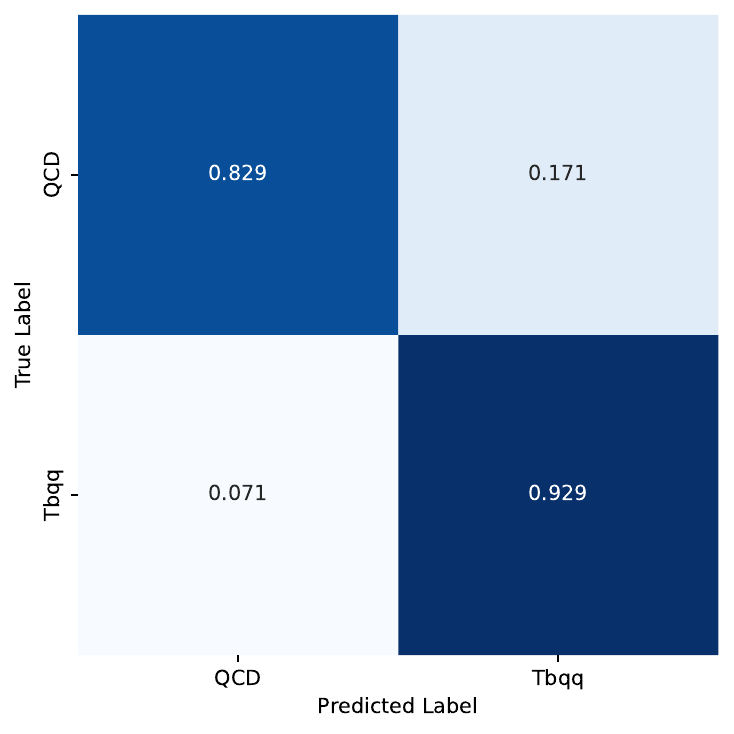}
    
    \includegraphics[width=0.48\linewidth]{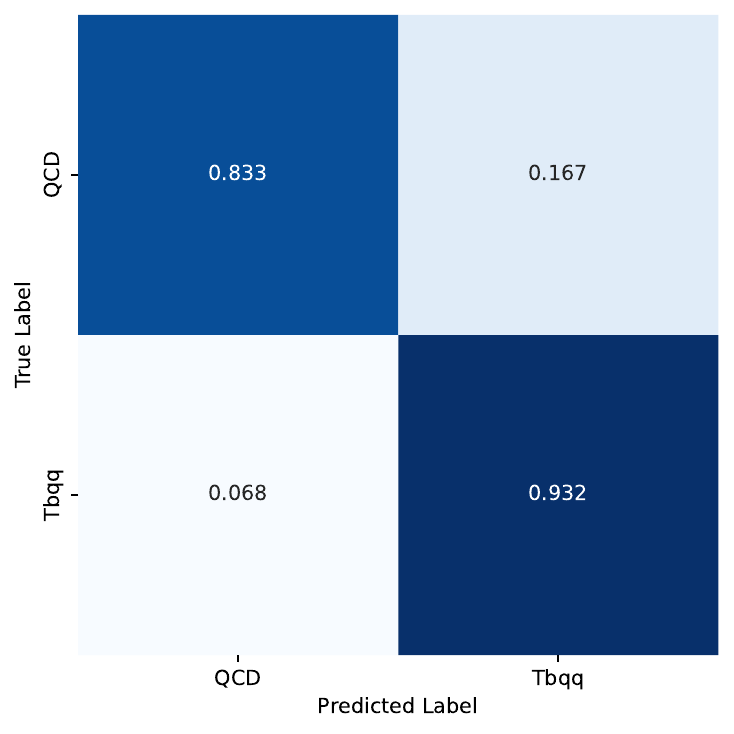}
    \includegraphics[width=0.48\linewidth]{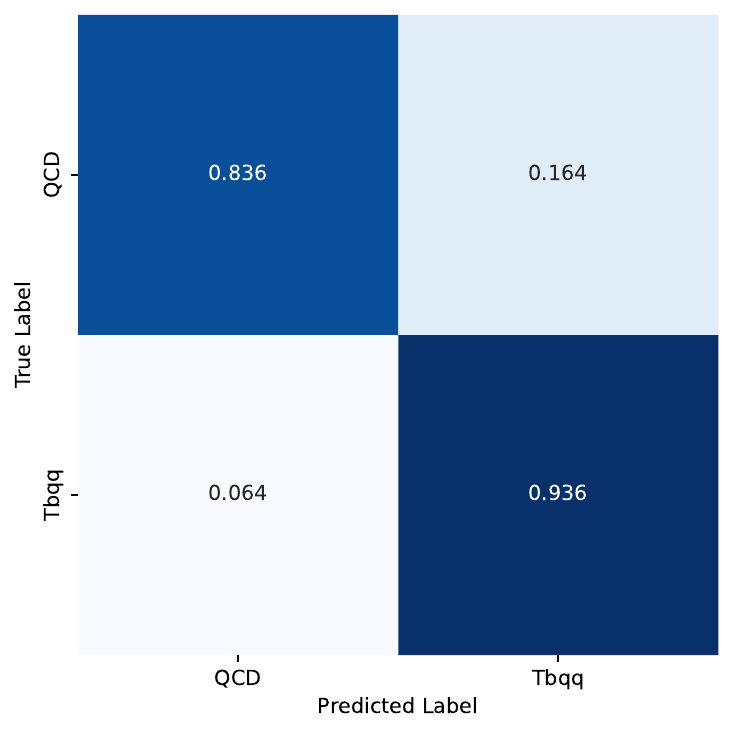}

    \caption{Confusion matrix of BDT for hadronic top class (\texttt{Tbqq}) vs QCD class from \jetclass on the learned jet embeddings from pretrained nano (top left), lite (top right), mini (bottom left), and base (bottom right) models.}
    
    \label{fig:confusion-matrix}
\end{figure}

\paragraph{Backbone Embeddings}
Figure~\ref{fig:embedding-PCA-tSNE-all} shows the learned jet embeddings from pretrained nano, lite, mini, and base models using PCA decomposition and t-SNE embedding. 
Figure~\ref{fig:confusion-matrix} presents the confusion matrices of BDT classifiers for hadronic top class vs QCD class trained on these learned embeddings, which demonstrates that all models achieve well-balanced classification performance, with the base and mini models showing slightly better true positive rates for top jet identification compared to the nano and lite models.
For t-SNE visualization, 10,000 randomly chosen jets are embedded and plotted.
For PCA corner plots, principal component analysis is performed on the entire test dataset, with 1,000 randomly selected jets used to create the scatter plots for off-diagonal elements.
PCA and t-SNE are implemented using the \textsc{scikit-learn} package~\cite{scikit-learn}, with visualizations created using the \textsc{Matplotlib} package~\cite{matplotlib} and array processing performed using the \textsc{NumPy} package~\cite{Numpy}.
The similarity of the PCA decompositions between the mini and base models explains their similar BDT accuracies.

\end{document}